# Engineering magnetic anisotropy and the surface of epitaxial Fe films using ion beam erosion; unveiling self-assembly and tunability

Anup Kumar Bera[1,2], Md. Shahid Jamal[1], Avinash Ganesh Khanderao[1,3] and Dileep Kumar[1,*]

[1]UGC-DAE Consortium for Scientific Research, Khandwa Road, Indore-452001, India
[2]Department of Materials Engineering, Indian Institute of Science, Bangalore-520016, India
[3]Department of Physics, Jagadamba Mahavidyalaya, Achalpur City-444806, India

[*] Corresponding author: dkumar@csr.res.in

The engineering of surface morphology and structure of the thin film is one of the essential technological assets for regulating the physical properties and functionalities of thin film-based devices. This study investigates the evolution of surface structure and magnetic anisotropy in epitaxially grown ultrathin Fe films on MgO (001) substrates subjected to multiple cycles of ion beam erosion (IBE) after growth. Ultrathin Fe film grows in 3D island mode and exhibits intrinsic fourfold magnetic anisotropy. After a few cycles of IBE, the film displays an induced uniaxial magnetic anisotropy that leads to a split in the hysteresis loop. In addition, clear and conclusive evidence of IBE mediated (2×2) reconstruction of the Fe surface has been observed. We also demonstrate that thermal annealing can reversibly tune the induced UMA and surface reconstruction. The feasibility of the IBE technique by properly selecting ion beam parameters for modification of surface structure has been highlighted apart from conventional methods of tailoring the morphology for tuning of UMA. Thus, the present work paves a way to explore the IBE-induced self-assembling phenomena further.

# Introduction

Magnetic thin films are an area of intense study that has expanded tremendously in the previous decade. The atomic-scale investigation of the link between ultrathin epitaxial film and magnetic properties is a challenging problem due to the presence of several effects, such as surface/interface morphology [1], interface diffusion [2] and hybridization [3], film structure [4], stain [5], etc. On the other hand, these dependencies open up the possibilities of the fine tailoring of magnetic properties according to the desired functionalities. Magnetic anisotropy is a crucial characteristic of ferromagnetic materials from a fundamental perspective to study magnetic switching and for numerous practical uses, such as magnetic memory devices [6]. The structure and morphology of ultrathin films directly affect their magnetic anisotropies. The fundamental processes behind magnetic anisotropies include dipolar and spin-orbit interactions [7]. The dimensionality and external shape of the particular ferromagnetic body and the local lattice symmetry are all important factors in magnetic anisotropy. Advances in sample preparation methods, the availability of ultra-high vacuum chambers, and the affordability of in-situ compatible atomic scale characterization tools have enabled the growth and characterization of ultrathin films. Low energy ion beam erosion (IBE) is a new, versatile and handy tool that has found its application for tailoring various physical properties in diverse fields such as photonics [8], transport [9], sensing [10], plasmonic [11], surface chemistry [12] bio-medical applications [13] etc. It is observed that low energy IBE of thin film and substrate surface create a wide range of morphology, including periodic ripple, nanoholes, arranged arrays of dots and pits [14][15][16] etc., depending upon the ion energy and angle of incidence. The periodic ripple-like pattern is formed due to competition between curvature-dependent ion beam sputtering that roughens the surface and ion or temperature-dependent viscous flow that smooths the surface [17]. Since the structure and morphology of ultrathin films directly affect their magnetic anisotropies, researchers are utilizing the IBE technique for fine and flexible tailoring of magnetic anisotropy.

The thin film deposited over the rippled substrate or direct IBE of the thin film was found to induce an additional component of UMA due to shape anisotropy in polycrystalline [18][19][20] and epitaxial thin film [21][22]. However, due to intrinsic crystalline anisotropy in the epitaxial thin film, a competition between shape and crystalline anisotropy is often observed [1]. This competition gives rise to the multistep switching process, splitting the hysteresis loop [23] etc. Therefore, investigation in this field is ongoing. Different phenomena emerge as a result of bombarding solid surfaces with energetic particles, all of which are strongly linked to the energy deposition processes of the incoming ions. It may be noted that the interaction of solid surfaces with low-energy ions can also cause mass redistribution when sputtering is negligible in the near-surface region [24][25][26]. Therefore, surface structure can also be altered with the low-energy ion beam. However, no studies have considered the corresponding ion beam interaction-induced structural changes in the polycrystalline and epitaxial thin film. In earlier studies, we have shown through in-situ investigation that the IBE of the polycrystalline thin film changes surfaces crystallographic texture [27][28]. This motivates us to further explore the effect of IBE on magnetic anisotropy and structural alternation of the epitaxial thin film. It may be noted that the reconstructed surface can control the growth of material by altering diffusion and nucleation, can produce a non-uniform interface that can affect the physical properties where the interface constitutes a large fraction of ultra-thin film, can induce uniaxial magnetic anisotropy[29]. Furthermore, surface properties like as chemical reactivity, energy, work function, vibration, and electronic state are all influenced by atomic geometry.

Fe/MgO system has recently received much attention due to its substantial interfacial PMA [30] and applicability in TMR devices [31]. For the present study, we have chosen Fe/MgO system for its easy growth and epitaxial relationship with the MgO substrate. Epitaxial Fe film was grown in the UHV chamber on a MgO substrate. The Fe film surface was etched by Ar ion in several steps, and the evolution of surface structure and magnetic anisotropy was studied in situ simultaneously. We observed that the film exhibits an additional UMA that causes a split in the hysteresis loop. Interestingly, a 2×2 surface reconstruction has been observed for the first time. These results offer a new dimension to the knowledge of self-assembled surface morphology development during IBE and propose that modelling studies be conducted to capture these events.

# Experimental Section.

Present experiments have been performed inside an ultra-high vacuum (UHV) chamber with a base pressure of ~$5\times10^{-10}$ mbar or better to avoid oxidation and absorption of contaminations on the substrate and film surface. The chamber is equipped with the facilities for thin film deposition by electron beam evaporation and in-situ characterization using MOKE and RHEED. It is also attached with an ion gun to clean and irradiate the sample surface. In the present study, a MgO(001) substrate is mounted on a sample holder attached to a heater assembly on a high-precision rotatable UHV-compatible manipulator. Cleaned and ordered substrate surface is prepared by cycles of Ar+ ion beam sputtering using ultra-high-purity 99.999% Ar+ gas and subsequent annealing at 600°C until a sharp RHEED pattern operated at an electron acceleration voltage of 20 keV and 1.45A emission current is observed. In order to prevent any growth-induced anisotropy in the Fe film, deposition is done normally to the substrate surface. Fe film thickness is monitored in-situ during deposition by a water-cooled calibrated quartz crystal monitor. After deposition, the film surface is eroded with Ar+ ion of energy 1keV at an angle 50º from the normal surface. Azimuthal angle-dependent MOKE hysteresis loop is recorded using a polarized He-Ne laser light (λ = 632.8 nm) and a UHV-compatible electromagnet in longitudinal geometry. To draw intercorrelation among structural, morphological and magnetic anisotropy in conjugation with the direction of IBE, MOKE and RHEED measurements were performed by rotating the sample with respect to the IBE direction. To further understand the interface magnetism (hyperfine field, spin orientation), a $^{57}Fe$ epitaxial thin film of relatively higher thickness, 25nm, was deposited on the same MgO(001) substrate. We performed synchrotron-based nuclear resonance scattering (NRS) measurements near the critical angle for depth-resolved magnetic properties. NRS measurements were performed at the nuclear resonance beamline P01 at PETRA III, DESY (Deutsches Elektronen-Synchrotron, Hamburg), using energy 14.4 keV, which corresponds to the $^{57}Fe$ Mossbauer transition to determine the magnitude and direction of magnetic hyperfine fields at the bulk and interface regions of Fe.

# Results and Discussions

Figure 1 shows the RHEED images of the MgO substrate taken along (a) $[110]_{MgO}$ and (b) $[100]_{Fe}$ direction after several cycles of the cleaning process. The presence of sharp streaks confirms the single crystalline nature of the substrate. Fe film of thickness 25Å is deposited on this cleaned substrate. The corresponding RHEED images of the Fe film surface are presented in Fig 1(c) and (d). The presence of streaks in the images indicates the epitaxial growth of Fe film with its well-established crystallographic orientations $(001)[100]_{Fe} \| (001)[110]_{MgO}$. It may be noted that the streaks are not continuous but somewhat spotty. This type of RHEED pattern is observed due to 3D diffraction from the island-like structures on the surface [32]. It indicates that the Fe film surface is not atomically flat, which may be due to the island form growth of Fe on MgO. A close inspection reveals that the spots in the RHEED pattern of Fig. 1(e) exhibit a centred square shape and in 1(f) displays a rectangular shape arrangement which reflects the symmetry of the $[100]_{Fe}$ and $[110]_{Fe}$ direction of Fe as it possesses bcc structure.

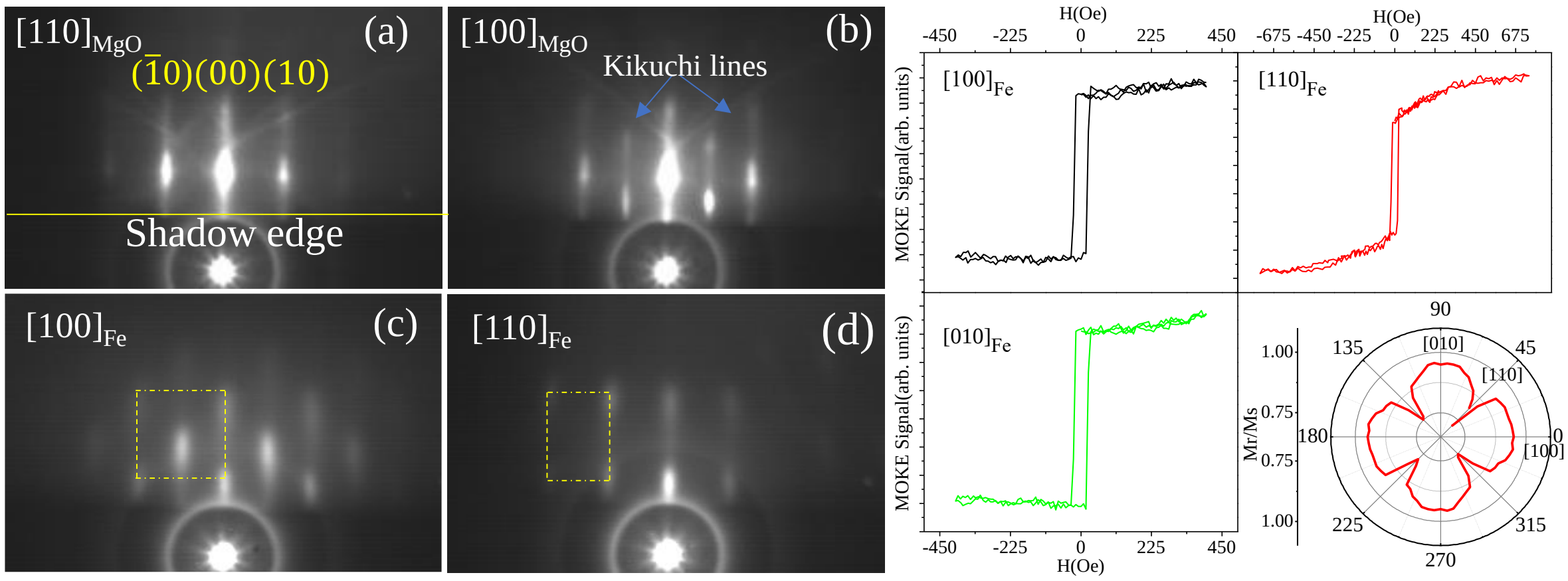

Figure 1: In-situ RHEED images of MgO (001) substrate taken along (a) [110] and (b) [100] directions. (c) and (d) represents RHEED images along [100] and [110] direction of 25Å thick Fe film epitaxially grown on the MgO substrate. In-situ MOKE hysteresis loop taken along [100], [110] and [010] direction of Fe. The corresponding polar plot of normalized remanence value.

The magnetic characterization of the sample was carried out in-situ by MOKE measurements in longitudinal geometry, and analysis was focused on the determination of strength as well as the orientation of the in-plane anisotropy axis. The MOKE hysteresis loop taken by applying magnetic field H along $[100]_{Fe}$, $[010]_{Fe}$ and $[110]_{Fe}$ directions are presented in Fig 2. It is observed that MOKE loops along [100] and [010] direction exhibits very high squareness with normalized magnetization (Mr/Ms) close to 1, whereas MOKE loop along the [110] direction have normalized magnetization close to 0.7(MsCos45º). The polar plot of the remanence (Mr) value normalized to its saturation (Ms) value is plotted in Fig. 2(d). The shape of the graph indicates that the film exhibits well-defined biaxial magnetic anisotropy, which is expected for the bcc Fe structure. The film surface was eroded with Ar ion step by step. Each step consists of 2.5min IBE by 1keV Ar ion. The projection of ions on the film surface was along $[100]_{Fe}$ direction, corresponding to the thermodynamically preferred step orientation direction. Their evolution was investigated as a function of IBE time to draw the correlation between surface structure and magnetic anisotropy.

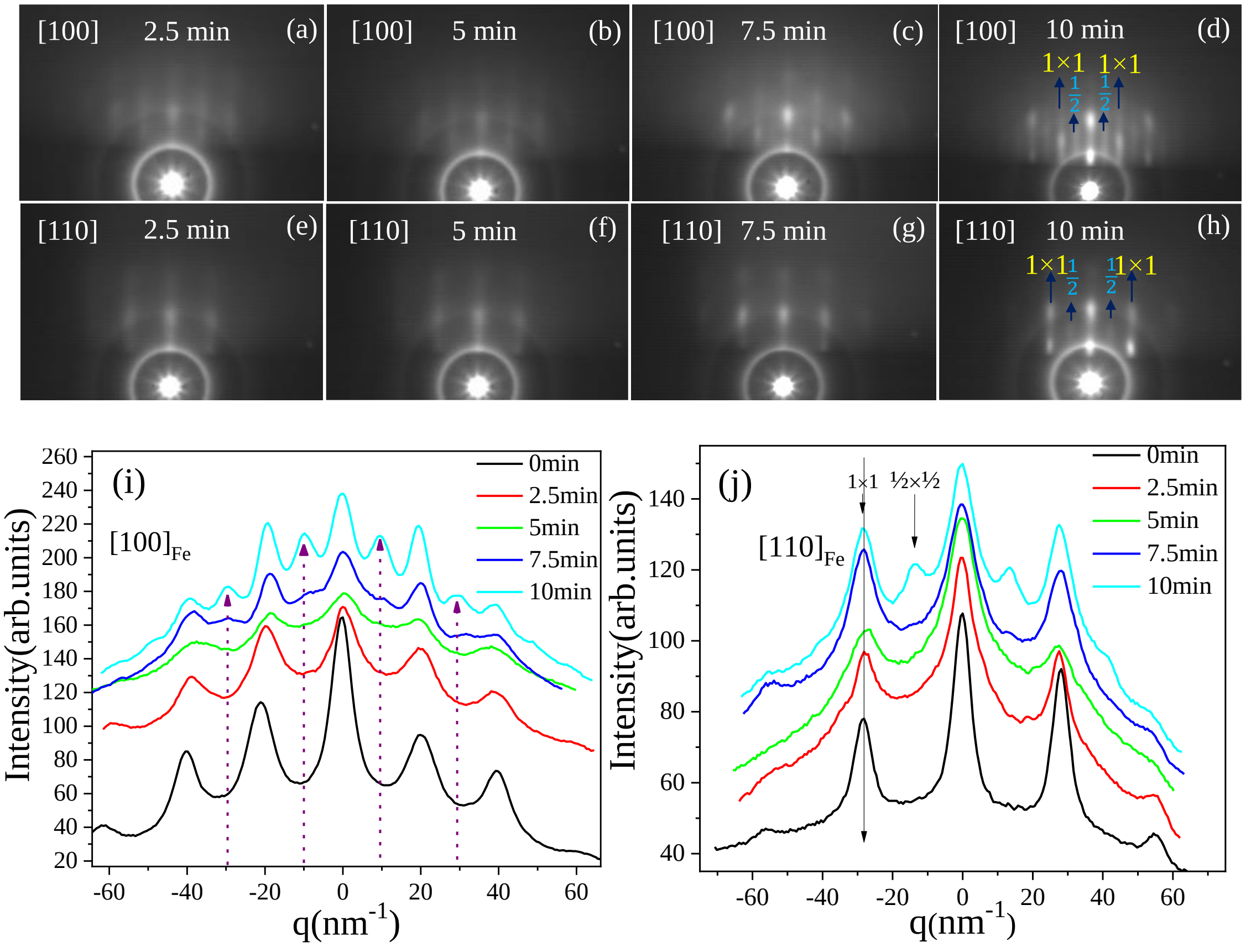

Fig 3: In-situ RHEED images taken along (a-d) $[100]_{Fe}$ direction, (e-h) $[110]_{Fe}$ direction after successive 2.5min ion beam erosion. Line profile of RHEED intensity distribution pattern after successive 2.5 min of IBE extracted from RHEED image taken along (i) $[100]_{Fe}$ and (j) $[110]_{Fe}$ direction.

The evolution of RHEED images taken along $[100]_{Fe}$ and $[110]_{Fe}$ direction at an interval of 2.5min IBE is presented in Fig 3. It is observed that after 7.5 min of IBE, an intermediate line in between integer streaks appears, which becomes more prominent after the 4$^{th}$ cycle (10 min) of IBE. For quantitative information about the position of new streaks, the line profiles have been extracted from the RHEED images as a function of IBE time and have been plotted in Fig. 3(i,j). We found that the new streaks are between the integer streaks. A similar diffraction pattern is also recorded along equivalent crystallographic directions, and they are essentially symmetric with respect to the central streak. It may be noted that the intensity of the new spots is much smaller compared to that of the primary structure. Therefore, it should originate from the emerging 2×2 structure. Furthermore, a slight difference between the lattice parameter has also been observed between the as-deposited and reconstructed states. The physical origin of surface reconstruction involves displacement of the surface atom, increment or decrement of surface density due to the addition or removal of the surface atom that leads to minimization of the number of surface atom dangling bonds, improvement of the compactness of outermost surface layer etc. [33][34][35]The interaction of energetic ions with solid surfaces causes erosion(sputtering) and displacement/ transportation of materials on the surface and near-surface regions. Noble gas ions with energies of about one keV have traditionally been used for experiments. In situations when the incoming ions' energies are just a few hundred electron volts or less, sputtering is almost nonexistent [24][36]. However, the formation of nanostructures in the form of ripples and disordered arrays of spikes has also been seen in this low-energy regime [24][37][38]. In this low-energy domain, mass redistribution (MR) caused by the inelastic displacement of atoms significantly increases due to the transfer of momentum from the incident ions to those atoms residing near the solid surface [25][26][39]. Based on ion energy and target material, hundreds of materials may be transported to another place without sputtering. The ion flux used for the present experiment ranges from $10^{11}$-$10^{12}$ ions/cm$^2$, approximately 3-5 orders of magnitude lesser than the conventional ion fluxes used for ordered patterned formation. Therefore, ion-induced mass redistribution by the exposure of material with low energy and low flux causes surface reconstruction.

The evolution of the hysteresis loop along ($\phi$=0°) and across ($\phi$=90°) to the IBE direction, i.e., along $[100]_{Fe}$ and $[010]_{Fe}$ direction after the 2$^{nd}$ and 4$^{th}$ cycles, is reported in Fig 4. One can observe that the height of the MOKE signal decreases with the IBE cycle. This is because, in the ultrathin regime, the height of the saturated MOKE signal is proportional to the film thickness. Therefore, this change can be rationally explained by considering the decrement in film thickness associated with the sputter removal of the film [27]. A closer inspection reveals that after 10min of IBE, the loop along [100] direction has split into two semi-loops. The magnetic remanence also decreases from ≈1 in the as-deposited state to ≈0.15 after the 4$^{th}$ cycle of IBE. However, hysteresis loops perpendicular to the IBE direction retain their shape as they were as prepared. It may be noted that both [100] and [010] directions are crystallographically and magnetically equivalent direction. Therefore, this change in the shape of the loop is associated with the change in the magnetization switching process due to the onset of an extra UMA[40–42].

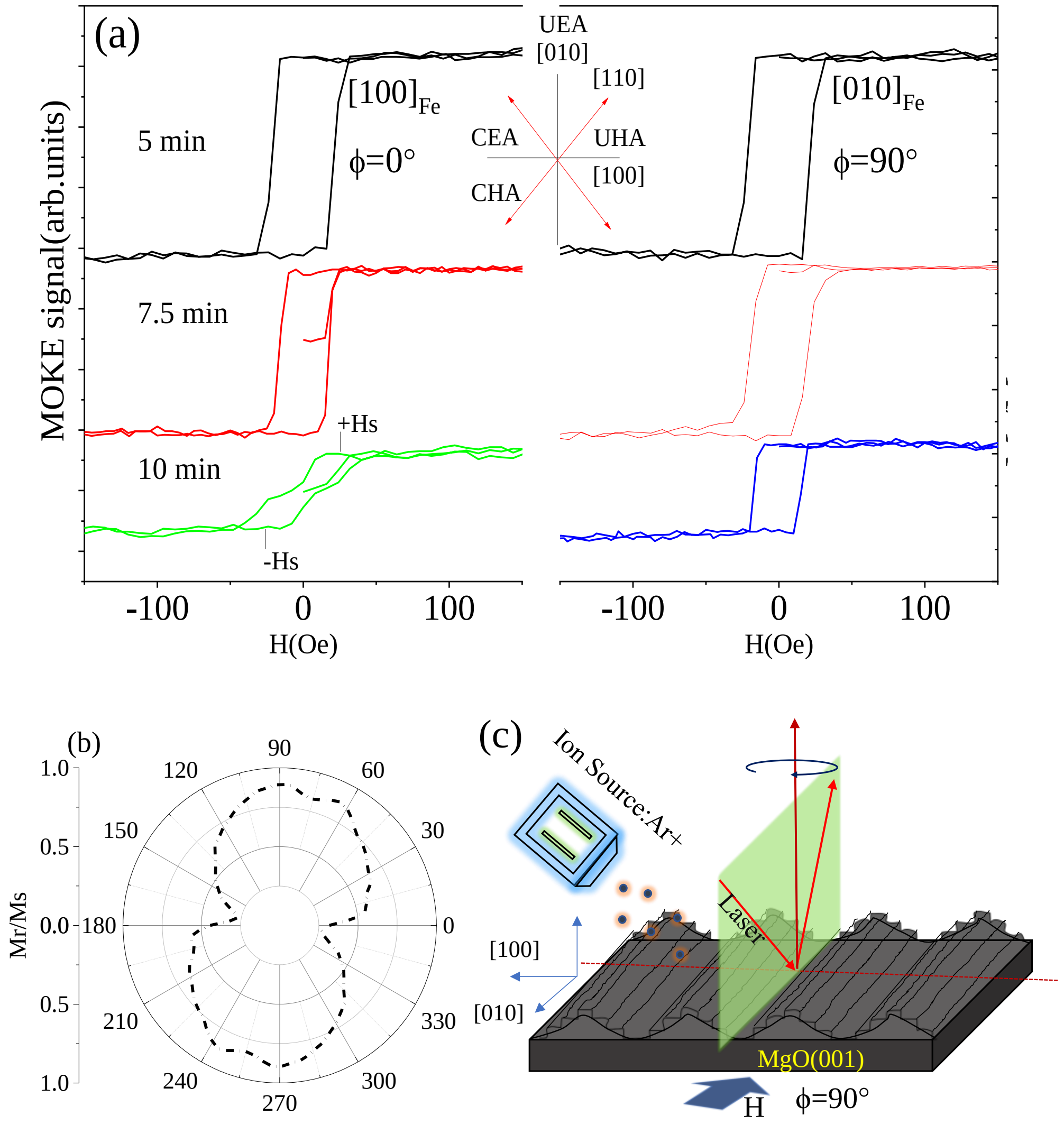

Fig 4: (a) Evolution of MOKE hysteresis loop along 0° and 90° with IBE cycle. (b) Polar plot of normalized remanence value after 4th cycle of IBE. (c) Schematic of IBE and crystallographic directions.

In the present case, UMA easy axis is oriented perpendicular to the IBE direction, i.e., along the [010] direction. Therefore, [010] direction is the easiest direction whereas [100] direction is an intermediate direction due to the presence of cubic easy and uniaxial hard axis. A quantitative measurement of the strength of induced UMA can be obtained by measuring the shift field Hs which is proportional to the split in the loop. It is defined as the difference between the zero fields and centre of one shifted loop, i.e., $Hs=(Hs_2 -Hs_1)/2$ as pointed out in Fig. and directly connected to UMA constant Ku=MsHs, where Ms=1737 emu/cm$^3$ is the saturation magnetization. Ku=1737*18=3.13*10$^4$ erg/cm$^3$. $K_1$= $(1737)^2/2s$= the initial linear slope of the loop[43][22]. The strength of the induced UMA, which is proportional to the shift in the semi-loop, is lower than the intrinsic cubic anisotropy.

The energy of anisotropic magnetization related to biaxial (Kc) and uniaxial (Ku) anisotropy constant via the relation: $E = \frac{K_C}{4}\mathrm{Sin}^2(2\varphi) + K_U\mathrm{Sin}^2(\varphi) - \boldsymbol{M}.\boldsymbol{H}$. Since the film is epitaxial in nature, an intrinsic MCA is present. The origin of induced UMA can be devoted to the MCA due to the unbalance between the number of steps oriented parallel and perpendicular to the IBE direction, low coordinated atoms present at step edge or surface defects sites as well as shape anisotropy due to the ripple structure itself[44–46]. The formation of unidirectional correlated morphology breaks the four-fold symmetry of the Fe surface and induces a UMA oriented parallel to the ripple direction. Since the ion sculpting process is mainly limited to the surface and subsurface layers, magnetic anisotropy with strength proportional to the inverse of film thickness is expected. The appearance of anisotropy Ku after 10 min of IBE results from the gradual buildup of the density of atomic steps along the $[100]_{Fe}$ directions, along with gradually increasing magnetostatic contributions[46]. To draw the correlation between surface structure on the induced uniaxial magnetic anisotropy, its structural evolution was investigated by RHEED as a function of the temperature. The RHEED images taken along $[100]_{Fe}$ direction at different temperature is presented in Fig 5. We observe that the other streaks start to disappear close to 300ºC and completely disappear at 320ºC. The line profile drawn at different temperatures is presented in the insets of each image. It also confirms that the intensity of the additional streaks gradually decreases with temperature and completely disappears at 320ºC. The MOKE hysteresis loop taken at 320ºC along ϕ=0º,45º and 90º directions are displayed in Fig 5.

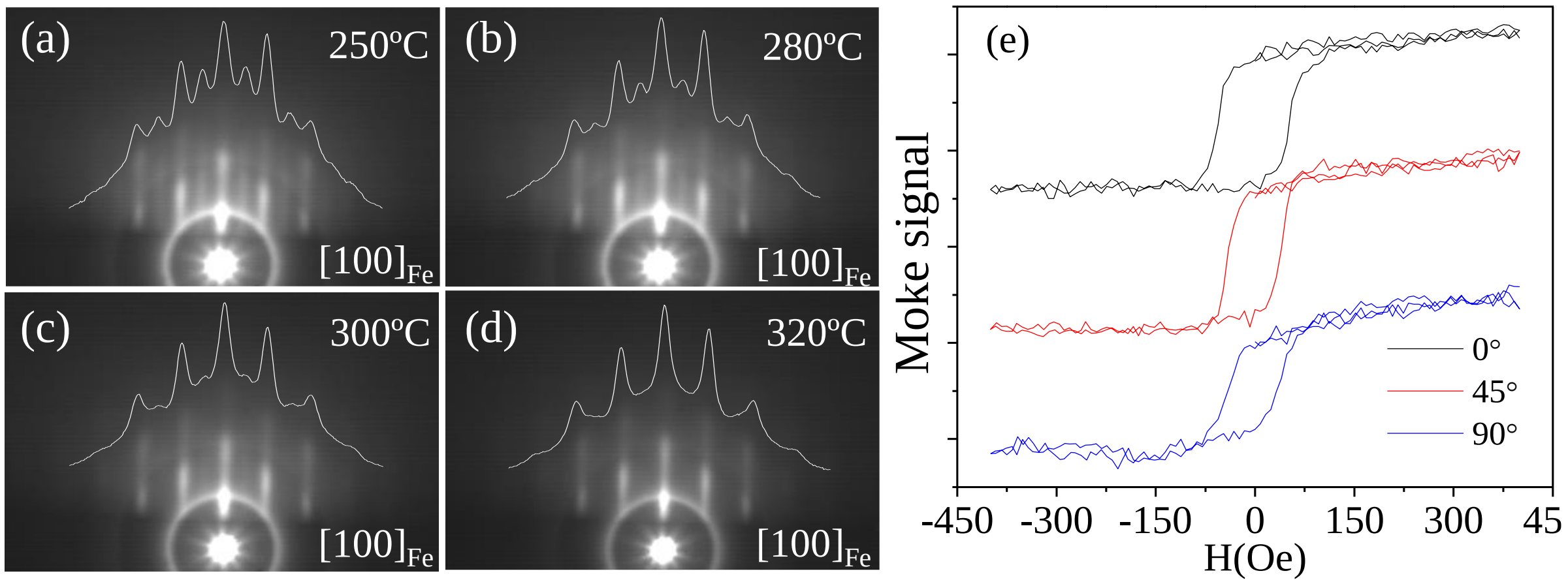

Fig 5: (a-d) temperature-dependent evolution of RHEED pattern along $[100]_{Fe}$ direction. The corresponding line profile of the RHEED intensity distribution pattern is also shown in the respective image. (e) In-situ MOKE hysteresis loop taken at room temperature after heating at 320ºC along different orientations w.r.t IBE direction.

We observe that all the loops exhibit the same magnetization-switching behaviour. It suggests that the induced UMA has disappeared. Intentionally, we have not capped the layer as it can alter the surface structure or magnetic property.

To further understand the interfacial magnetism (hyperfine field, spin orientation) due to surface reconstruction, a $^{57}$Fe epitaxial thin film of relatively higher thickness, 28nm, was deposited on the same MgO(001) substrate. The film surface was eroded with identical ion fluence and ion energy with which Fe thin film of 2.5nm thickness was eroded. The corresponding MOKE hysteresis loop and RHEED images taken before and after IBE is presented in the supplementary material. Similar to earlier studies, a 2×2 surface reconstruction was also observed.

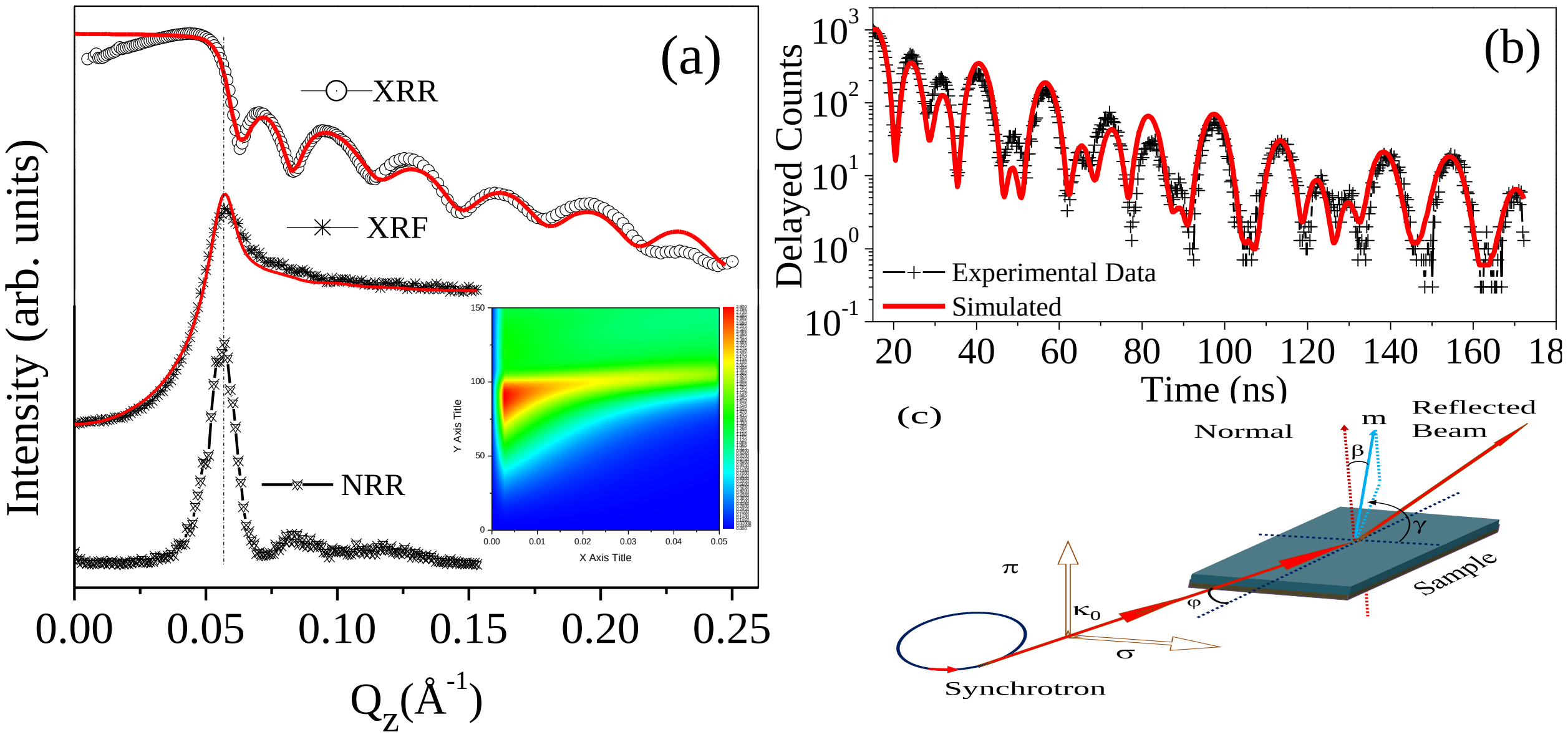

**Figure 6:** (a) XRR, XRF and NRR pattern of epitaxial $^{57}$Fe thin film on MgO (001) substrate. (b) Represents the corresponding nuclear forward scattering time spectra. The time spectra are fitted using REFTIM software. (c) Schematic of the experimental set-up.

The distribution of electric field intensity (EFI) of X-ray within the layer as a function of the scattering vector q (incident angle) is extracted using Parratt's approach, and the results are shown in Fig. 6(a). Around the angle of incidence = 0.224° (q = 0.057$Å^{-1}$), strong confinement of the EFI is observed. It indicates that the grazing incidence NRS (GINRS) measurement performed at q= 0.057$Å^{-1}$ will enhance the nuclear resonance counts. To validate it experimentally, nuclear resonance reflected intensity is collected as a function of q, together with XRR, and is displayed in Fig. 6(a). At q=0.057 $Å^{-1}$, the resonant intensity was higher than an order of magnitude. Considering the above fact, all GI-NRS data (time spectra) were collected at an incidence angle of q= 0.057 $Å^{-1}$. The measured GINRS time spectra are shown in Figure 6(b). Fig. 6(c) depicts a schematic illustration of sample orientation relative to the x-ray beam. To extract magnetic information, the observed time spectra were fitted using the REFTIM software with parameters obtained from simultaneous fitting of XRR, NRR, x-ray fluorescence (XRF). The best match to the time spectra was achieved by splitting the $^{57}$Fe layer into two layers with two overlapping hyperfine fields ($B_{hf}$) components, namely 32.7 T, which matches with bulk Fe, and a reduced component of 3T, representing the non-magnetic component. Interestingly the non-magnetic component is associated with out-of-plane 40° Fe moments orientations, whereas the Fe moments correspond to a 32T hyperfine field aligned in the plane of the film. The time evolution of forward nuclear scattering is determined by the interference of transitions determined by the hyperfine splitting of the nuclear levels. It should be noted that with cubic symmetry, the electric field gradient (EFG) and, hence, quadrupole splitting (QS) are likely to be absent[3]. However, the best-fitting parameters yield an EFG value of 0.978 mm/s for the layer corresponding to 3T. As a result, a finite QS value indicates a deformation from the cubic structure, as observed earlier from the change in the streaks spacing of the RHEED image.

## Conclusions

In conclusion, our study establishes a comprehensive understanding of the effects of IBE on ultrathin epitaxial Fe films. We successfully demonstrated the ability to induce uniaxial magnetic anisotropy in ultrathin Fe films through controlled ion beam erosion. The erosion process causes the formation of correlated nanoscale features, which in turn influences the film's magnetic behavior. Furthermore, our investigation revealed for the first time that IBE alters the magnetic properties and induces surface reconstruction in the Fe film. Thus, the present study shows that IBE holds great promise to engineer the morphology, structure and magnetic properties of thin films simultaneously. In-situ compatible surface-sensitive techniques such as RHEED must be utilized to further explore structural details and resolve questions related to self-assembled nanostructures, as demonstrated in this study. Notably, the induced uniaxial magnetic anisotropy and surface reconstruction hold great promise for various technological applications in thin film-based nanodevices. Additionally, controlled surface reconstruction can be utilized to enhance the performance of magnetic devices through improved surface-to-volume ratios, increased interfacial interactions, and tailored magnetic domain structures. Future research can explore the optimization of ion beam parameters, investigate the dynamics of surface reconstruction, and explore the feasibility of integrating these findings into practical applications.

**Acknowledgement**

The authors thank Dr. H. C. Wille and Dr. Olaf Leupold for the NFS measurement at the P01 beamline, PETRA III, Desy, Germany. The authors also acknowledge the Department of Science and Technology (Government of India) for providing financial support within the framework of the India@DESY collaboration.